\documentclass[aps,amsmath,amssymb,twocolumn,superscriptaddress]{revtex4}%
\usepackage{amsmath}
\usepackage{xcolor}
\usepackage{graphicx}
\usepackage{dcolumn}
\usepackage{bm}
\usepackage{float}
\usepackage{epsfig}
\usepackage{amsfonts}%
\usepackage{color}
\usepackage[utf8]{inputenc}
\usepackage{amssymb}
\usepackage{natbib}
\usepackage{soul}
\usepackage{scalerel}
\usepackage{cancel}
\usepackage{appendix}

\begin{document}

\title{Hybrid patterns and solitonic frequency combs in non-Hermitian Kerr Cavities}
\author{Salim B. Ivars}
\affiliation{Departament de Física, Universitat Politècnica de Catalunya (UPC), Rambla Sant Nebridi 22, 08222, Terrassa, Barcelona, Catalonia, Spain}

\author{Carles Mili\'{a}n}
\affiliation{Institut Universitari de Matem\`{a}tica Pura i Aplicada, Universitat Polit\`{e}cnica de Val\`{e}ncia, 46022 Val\`{e}ncia, Spain}

\author{Muriel Botey}
\affiliation{Departament de Física, Universitat Politècnica de Catalunya (UPC), Rambla Sant Nebridi 22, 08222, Terrassa, Barcelona, Catalonia, Spain}
\author{Ramon Herrero}
\affiliation{Departament de Física, Universitat Politècnica de Catalunya (UPC), Rambla Sant Nebridi 22, 08222, Terrassa, Barcelona, Catalonia, Spain}

\author{Kestutis Staliunas}
\affiliation{Departament de Física, Universitat Politècnica de Catalunya (UPC), Rambla Sant Nebridi 22, 08222, Terrassa, Barcelona, Catalonia, Spain}
\affiliation{Institució Catalana de Recerca i Estudis Avançats (ICREA), Passeig Lluís Companys 23, E-08010, Barcelona, Spain}
\affiliation{Vilnius University, Faculty of Physics, Laser Research Center, Sauletekio Ave. 10,  Vilnius, Lithuania}

%
\begin{abstract}

We unveil a new scenario for the formation of dissipative localised structures in nonlinear systems. Commonly, the formation of such structures arises from the connection of a homogeneous steady state with either another homogeneous solution or a pattern. Both scenarios, typically found in cavities with normal and anomalous dispersion, respectively, exhibit unique fingerprints and particular features that characterise their behaviour. However, we show that the introduction of a periodic non-Hermitian modulation in Kerr cavities hybridises the two established soliton formation mechanisms, embodying the particular fingerprints of both. In the resulting novel scenario, the stationary states acquire a dual behaviour, playing the role that was unambiguously attributed to either homogeneous states or patterns. These fundamental findings have profound practical implications for frequency comb generation, introducing unprecedented reversible mechanisms for real-time manipulation.

\end{abstract}

\maketitle

Pattern formation in extended systems is a far-from-equilibrium phenomenon that rules the dynamics of nonlinear physical, chemical or biological systems \cite{cross1993pattern,arecchi1999pattern,staliunas2003transverse,lugiato2015nonlinear,turing1990chemical,knobloch2015spatial}. The existence and stability of pattern states (PS), together with that of the so-called flat or homogeneous states (HS), lies at the core of the dissipative localised structure (LS) formation \cite{schapers2000interaction,meron1992pattern,pomeau1986front}, including optical solitons and frequency combs. The emergence of solitons in different regimes of nonlinear systems is generally associated with the existence of two stable solutions, either a PS and HS for anomalous group velocity dispersion (GVD) or two stable HSs for normal GVD. {Some non-linear systems have additionally shown to hold tristability: three different stable HS \cite{dumeige2011stability,anderson2017coexistence,milian2018cavity}, two PS and one HS \cite{tlidi1993transverse,spinelli1998spatial} or two HS and one PS \cite{zelnik2018implications,al2021unified,parra2022organization}. However, the two paradigms of formation of LS do not coexist and the character of the basal states is unique and well-defined.
}

In this letter, we unveil a novel hybrid scenario in the formation of solitons and patterns opening a new avenue in the comprehension of nonlinear systems. Specifically, we demonstrate that the introduction of a {periodic} modulation in Kerr cavities with normal GVD is qualitatively similar to the {effect of} Turing instability of the anomalous GVD, in the sense that the modulation introduces low amplitude Turing-like (periodic) states {without destabilising the system}. This is the key for the dual and hybridised mechanism that we demonstrate below, which is understood as the blending of the two traditionally separated LS forming regimes. As a result, new families of stable solitons, molecules, and patterns emerge in the normal GVD regime.

In the last two decades, the damped-driven Nonlinear Schrödinger Equation (NSE) has attracted significant attention \cite{kaup1978solitons,barashenkov1996existence,chembo2013spatiotemporal,lugiato1987spatial,haelterman1992dissipative}, for the mastering of microcavities and frequency comb generation\cite{bao2020turing,brasch2016photonic,kippenberg2011microresonator}. This model has been demonstrated to accurately describe the evolution of the light field in Kerr resonators. It {admits} LSs in the two canonical formation regimes associated with normal and anomalous GVD \cite{matsko2012normal,barashenkov1996existence}.

The introduction of inhomogeneities in the damped-driven NSE has been studied through the modulation of external injection or intracavity modulation \cite{copie2016competing}. In the first case, the soliton dynamics has been studied for temporal modulations of the driving field \cite{todd2023dynamics,hendry2018spontaneous}, { and the introduction of a spatial profile in the injected field} has shown to stabilise LSs \cite{tabbert2019stabilization}. In the second case, phase modulations by Electro-Optical Modulators within the ring cavity have shown to support synthetic dimensions \cite{tusnin2020nonlinear,englebert2023bloch,dutt2020single} and stabilisation of 3D solitons \cite{sun2023robust}. 

\begin{figure}
    \centering
    \includegraphics[width=0.86\columnwidth]{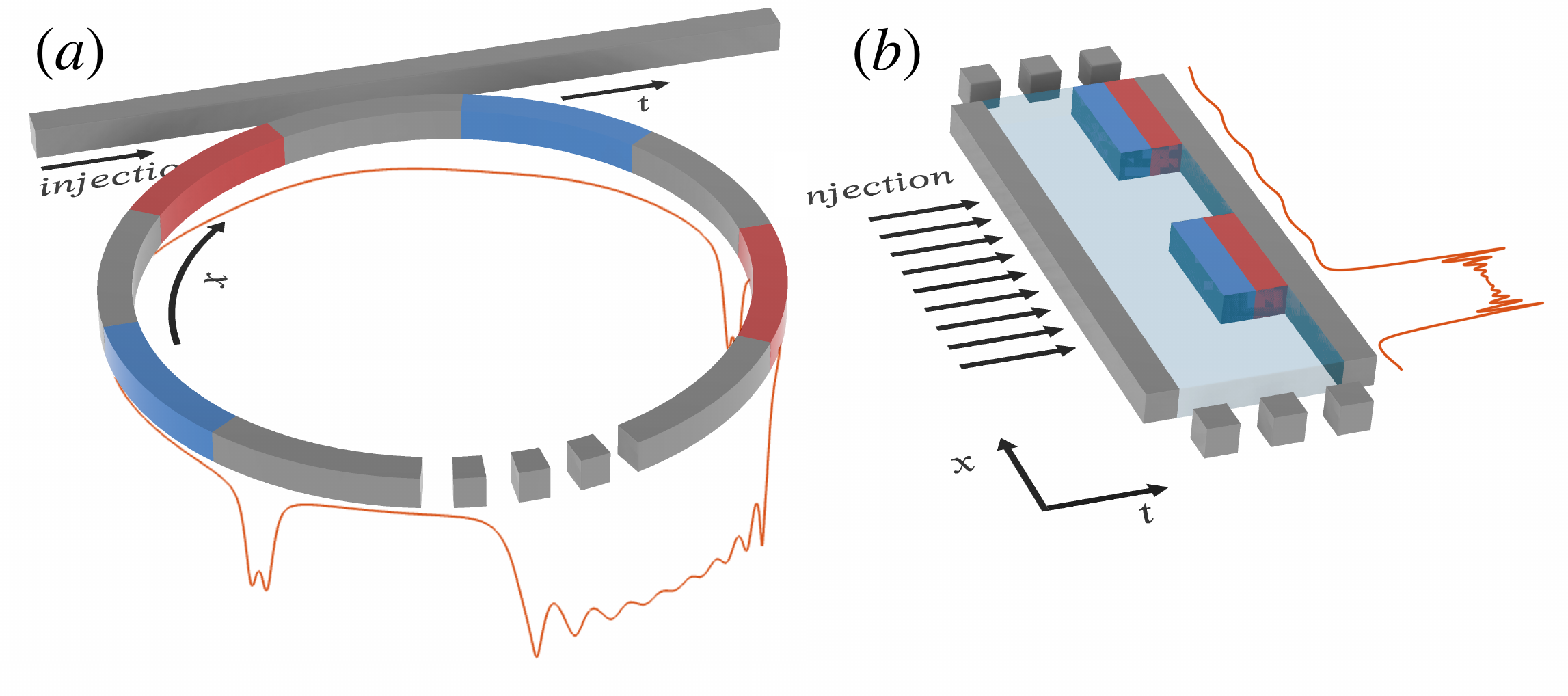}
     \vspace{-10pt}
    \caption{Schematic representation of possible experimental systems holding hybridised LSs formation. (a) {Kerr optical ring cavity with integrated electro-optical modulators for phase and amplitude modulation}. (b) Thin Fabry-Perot cavity with $\chi^{(3)}$ medium with loss and phase masks.}
    \label{fig:1}
\end{figure}

In parallel, the introduction of non-Hermitian (complex) potentials in non-linear systems has demonstrated the ability to induce a wide range of intriguing properties: unidirectional couplings arising from potential asymmetries \cite{ivars2022non,akhter2023non},  stabilisation of new solutions \cite{ivars2023stabilisation,ahmed2018stabilization}, the support of constant intensity waves \cite{makris2015constant}, the occurrence of exceptional points and jamming anomaly \cite{barashenkov2016jamming} or selective single-mode lasing \cite{feng2014single}, among others (see, \textit{e.g.}, \cite{ashida2020non,konotop2016nonlinear,suchkov2016nonlinear} for reviews). In optics, non-Hermitian potentials have been successfully implemented in recent experiments, \cite{ozdemir2019parity,cheng2023multi}. For instance, EOM have shown to effectively modulate the complex refractive index with integrated technology (see Fig.~\ref{fig:1}(a) {and \cite{hu2024integrated} for a recent review, and references therein}). In thin-film Fabry-Perot resonator \cite{lugiato2010difference,hachair2004cavity,brambilla1996interaction,peschel2003spatial,brambilla1997spatial}, the refractive index and loss modulations can be introduced with spatial phase and loss masks (see Fig.~\ref{fig:1}(b)). These cavities have emerged as attractive options for generating frequency combs \cite{cole2018theory,wildi2023dissipative}. Despite obvious differences with rings, the nonlinear stationary solutions of both systems share close similarities \cite{cole2018theory}.

The damped-driven NSE modelling passive Kerr dispersive cavities \cite{chembo2013spatiotemporal,lugiato1987spatial,haelterman1992dissipative} can be easily generalised to include a complex potential:
\begin{multline}
\label{eq:1}
\partial_t\psi=-i\partial_x^2\psi-(1+\delta i)\psi+2i\left|\psi\right|^2\psi+V{\left(x\right)}\psi+h,\\
            V{\left(x\right)}=me^{-i\phi}\cos{\left(\frac{2\pi px}{L}\right)},
\end{multline}
see \cite{tusnin2020nonlinear,englebert2023bloch,sun2023robust,dutt2020single} for the derivation of the normalised model and estimation of physical parameters. The cavity-laser detuning $\delta$ is our control parameter and the equation includes a third order Kerr nonlinearity, an external energy injection $h$ and a complex potential $V(x)$. Here, $m$ represents the depth of modulation, $p$ is the number of periods of length $L$ along the cavity path and $\phi$ represents the ratio between real and imaginary parts. In the current study, we fix the number of oscillations to $7$ without any loss of generality. For details on numerical methods see, \textit{e.g.} \cite{hult2007fourth,milian2018clusters}.

\begin{figure*}
\includegraphics[width=.95\textwidth]{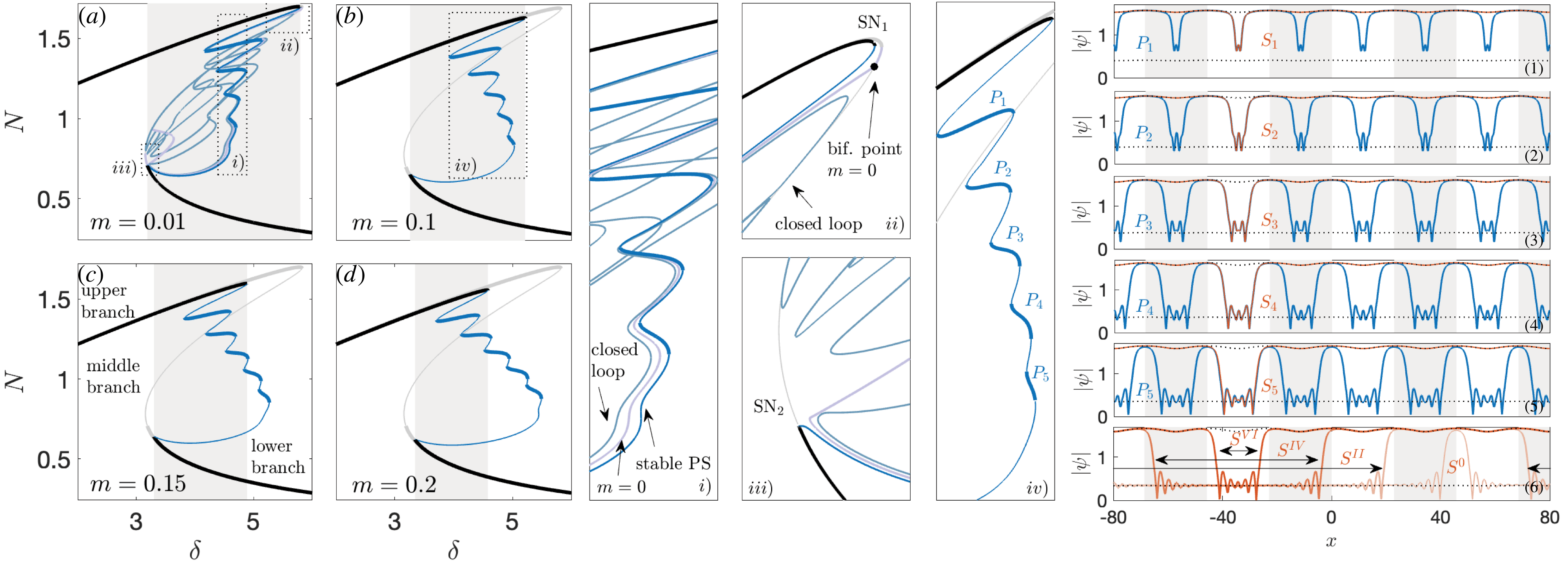}
 \vspace{-10pt}
    \caption{Bifurcation diagrams of HSs and PSs and corresponding spatial profiles. (a)-(d) Norm versus detuning, for increasing values of $m$, showing the HSs (black) and  PSs (blue and turquoise).The thick/thin curves indicate stable/unstable solutions, grey shaded area indicates HS bistability, and the HS (grey) for $m=0$ is depicted as a reference. Insets $i)$ and $ii)$ zoom in the overlap of the unmodulated PS branch (violet), the closed loop (turquoise) evolving from the middle branch, and the stabilised PS (blue); $ii)$ also shows the SN$_1$, and the PS bifurcation point for the unmodulated system. Inset $iii)$ shows the detachment of solutions from SN$_2$. Inset $iv)$ highlights the different families of stable PS, labelled as $P_n$, shown on the right panels along with solitons $S_n$. The bottom right panel illustrates even solitons, $S^{0-VI}$, with widths surpassing a single period $L$; periods are shaded in grey and white. $h = 1.7$.}
    \label{fig:2}
\end{figure*}

Regarding the formation of LSs, for the homogeneous bistability case (normal GVD), it is well-established that fronts or connections between two HS typically exhibit one exponential decaying oscillatory tail, and one monotonous connection related to the complex and real eigenvalues associated with each stable state \cite{ma2010defect,parra2016dark}. Such an oscillatory tail serves as a potential enabling two fronts to pin to each other forming a LS. When two fronts are close, the large oscillation amplitude leads to a strong effective potential generating a large coupling interval as a function of the control parameter. Conversely, as the distance between the fronts increases, this interval narrows. Ultimately, the LS can only form when the fronts have null relative velocity, a condition that only occurs for particular values of the parameters (at the Maxwell-Point). The above ingredients induce a collapsed configuration in the bifurcation structure known as \textit{collapsed snaking} \cite{ma2010defect}. For the bistability of a PS and a HS (anomalous GVD), the pinning of the two fronts, heteroclinic connections between both states, does not depend on the number of oscillations of the LS close to the PS but on the PS itself. Therefore, the generated branches associated with different LSs span the same interval of the control parameter, leading to what is known as \textit{standard homoclinic snaking}. The invariance of the system by a space-reversal symmetry, $x\rightarrow -x$, splits the LSs into two disjointed sets with even and odd number of oscillations of the pattern, respectively \cite{avitabile2010snake,burke2007homoclinic,tlidi1994localized}.In general, the two snaking diagrams are found in different regimes due to the distinct involved bistable states. These building mechanisms do not coexist, and thus collapsed and standard snake branches are not connected.

Therefore, a natural way to observe the hybridisation of the system is by inspecting the bifurcation diagram of these solutions. The study of the curve connected to the homogeneous solutions is summarised in Fig.~\ref{fig:2}. Importantly, in normal GVD regime, the stability of the upper homogeneous state is lost by a Saddle Node (SN) bifurcation. This is opposed to the anomalous GVD where the stability of the HS is lost at a TI point, arising due to the parametric four-photon mixing. The introduction of the periodic potential takes the role of Turing instability to modulate the system HSs. This induces the system to fold the HS curve, and connect the upper and lower HS with periodic solutions that become stable, see Fig.~\ref{fig:2}(a). The inset $i)$ zooms in the branch of the unstable PS for $m=0$, which emerges from the middle HS, see inset $ii)$. With the potential, this branch splits into two different solutions: one of them being stable PS -now connected to the HSs-, and the other unstable - and now part of a closed loop, that derives from the modulation of the middle branch-. Note that the branches of the two (stable/unstable) modulated and the unmodulated patterns are almost overlapping. {Added to this and shown on the insets $ii)$ and $iii)$, the middle HS becomes detached from SN$_{1,2}$ and forms an isola or closed loop.}

Since the PSs can be understood as a train of solitons, accordingly the branches follow the expected collapsed organisation until the width of a single soliton reaches the width of the period of the modulation. {This organises the discrete periodic structures, $P_n$, where the label $n$ refers to the position in the collapsed snaking.} A crucial effect of the introduction of the potential is the fact that the snaking structure gains a slant. This tilt is a finite-size effect, induced by the extra boundary introduced by the potential at each period \cite{kozyreff2009influence}. Such tilting grows with $m$, {eventually exceeding} the HS bistable region, see Fig.~\ref{fig:2}(a)-(d). {The non-Hermiticity of the potential favours this tilt, see the Supplemental material}. This brings to light the hybridisation and dual character of the system. Parts of the periodic solutions can no longer be understood as a product of the HS bistability but achieve the pattern character. The different families of solutions are displayed in the panels of Fig.~\ref{fig:2} and inset $iv)$, {where $S_n$ refers to a single soliton from a particular $P_n$ solution. Whenever the width of the soliton exceeds one period of the modulation, we denote the soliton as {$S^Q$}, being {$Q$} the number of full oscillations of the upper HS. Importantly, the modulated upper solution {now} takes the role of a pattern in the formation of LS.} Panel $(6)$ in Fig.~\ref{fig:2} shows solutions with an even number of oscillations, $S^{VI-0}$. 

\begin{figure*}
\includegraphics[width=.95\textwidth]{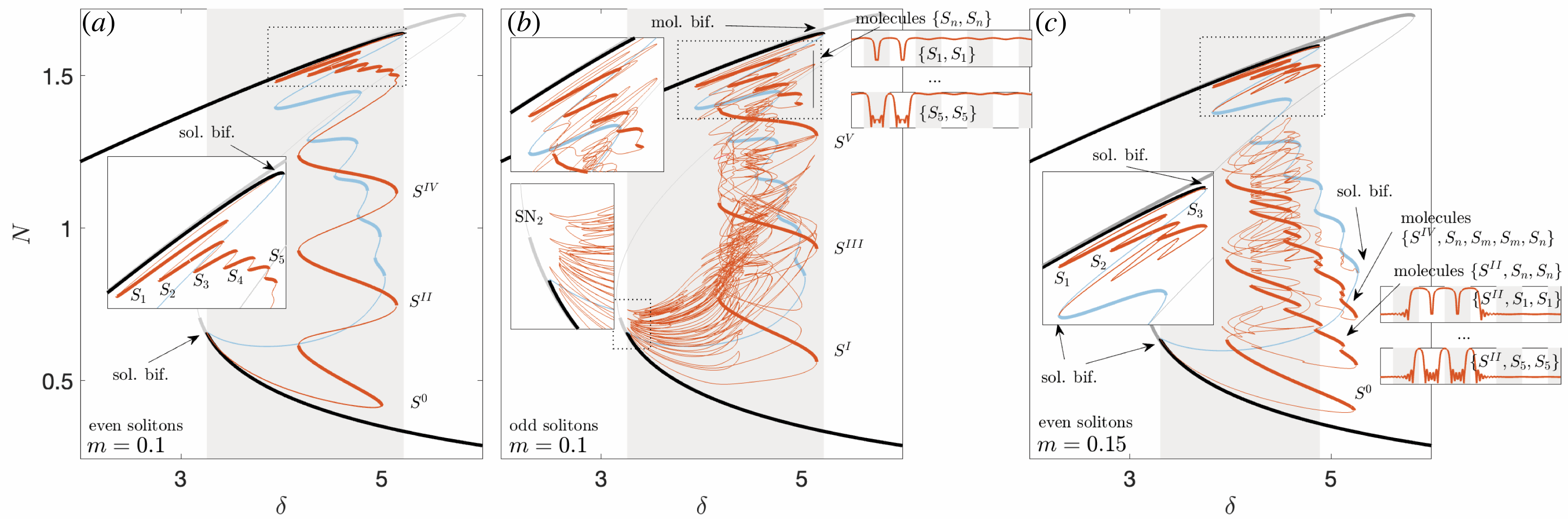}
 \vspace{-10pt}
    \caption{Bifurcation structure of LS in different hybridised scenarios. Norm versus detuning diagram of solitonic solutions (orange). (a) Family of even solitons for $m=0.1$ holding a tilted collapsed snaking followed by a standard homoclinic bifurcation structure. The inset zooms in the narrower soliton states $S_n$. (b) Family of odd solitons for $m=0.1$ with an analogous structure but with a collapsed snaking structure of two solitons molecule states of the type $\{S_n,S_n\}$, which profiles are provided on the panels. Lower-left inset: magnification of the SN$_2$ of the lower homogeneous solution. (c) Splitting of the even soliton solutions branch. Molecules of the type $\{S^{IV},S_n,S_m,S_m,S_n\}$ and $\{S^{II},S_n,S_n\}$, with field profiles provided on the right-hand panels. Inset: zoom in on the collapsed snaking structure of narrow solitons. In all plots, the thick/thin lines indicate stable/unstable states. Periodic solutions of Fig. \ref{fig:2} are included for completeness.}
    \label{fig:3}
\end{figure*}

In this study, multistability of two background solutions and patterns is achieved by the introduction of a modulation. Note that while such background solutions are not strictly homogeneous, the effect of the potential introduces relatively weak modulations, {see the dotted curves on panels $(1)-(6)$ in Fig.\ref{fig:2}}. With the potential, a new hybrid scheme where the basal solutions acquire a dual character is generated and summarised in Fig.~\ref{fig:3}. The figure captures the complexity and richness of the modulated system, where the typical collapsed and standard homoclinic snaking link. We differentiate two situations depending whether the PS existence lies only within or exceeds the HS bistable area. In the first case, depicted in Figs.~\ref{fig:3}(a) and \ref{fig:3}(b) for $m=0.1$, the LS branches containing the different families, bifurcates from the two main SNs. In Fig.~\ref{fig:3}(a) we show the branches of $S_n$. As for stabilised  $P_n$, their existence is organised in a tilted collapsed snaking, until its width exceeds a modulation period. Inspecting this structure, one could conclude that switching waves connecting both quasi-homogeneous states are the building blocks of these solutions. Yet, once the $P_n$ existence limit is exceeded, the homogeneous upper solution changes its role to a pattern. Broader solitons with even number of oscillations of the upper state, $S^{0,II,IV}$, are now organised in a standard homoclinic snaking. {This transition happens due to the duality of the upper HS induced by the potential}. In turn, Fig.~\ref{fig:3}(b) provides the analogous situation for even solitons, $S^{I,III,V}$, also organised in an intertwined homoclinic snaking. This existence (and detachment) of the branches of even an odd solitons highlights that the physics from the case of anomalous GVD is also present in the system. Interestingly, this hybridisation enlarges the stability area of bright solitons (for instance $S^{III-0}$). For no modulation, and without higher order terms, they are found just for the parameters of the Maxwell Point \cite{tlidi2010high,parra2017coexistence}. Notice, on the lower inset, the branch approaching the SN$_2$, and its difficult connection due to the symmetry of the system. At the top, we show how the odd soliton branch connects to a collapsed snaking structure of a molecule corresponding to two solitons in two neighbouring periods $\{S_n,S_n\}$, see the inset and the profiles on the corresponding panels. Solitons can now bond due to the oscillation of the upper state, induced by the potential. {Although not shown, stable molecules with different number of solitons, not necessarily laying in neighbouring periods exist and are stable.}

\begin{figure*}
\includegraphics[width=0.95\textwidth]{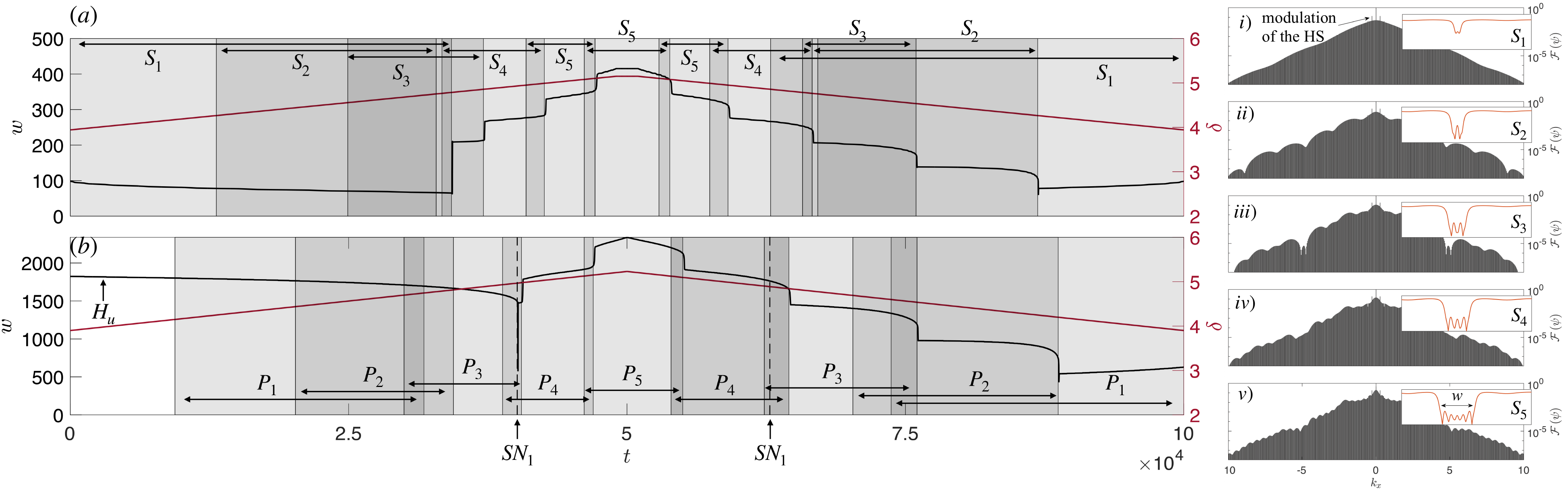}
 \vspace{-10pt}
    \caption{Temporal manipulation of frequency combs. Dynamical change of width, $w$, (black curves) of the propagating field and change of the detuning (red curves), for a) deterministic transitions between different families of Solitons $S_n$; b) turning on the system and switching between different patterns $P_n$. Shaded areas correspond to stable areas of the different states. Panels $\left(i\right)$-$\left(v\right)$ depict stable switchable comb spectra associated to solitons $S_{1-5}$.}
    \label{fig:4}
\end{figure*} 

An example of a case where the PS exceeds the bistable area is illustrated in Fig.~\ref{fig:3}(c). As commented earlier, this introduces unequivocally the pattern character to these periodic solutions. In this case, for solitons with even number of oscillations of the upper state, it is interesting to note two different organisations of the bifurcation structures. The previous branch splits. The collapsed snaking structure of $S_n$ remains, although connecting the upper HS to a PS, see inset. In turn, for broader solitons we can observe a reminiscence of the standard homoclinic snaking surpassing the HS bistable area, see $S_0$. Moreover, multiple branches of molecule-type solution develop within the snaking. As an example, these solutions associated with $S_0$ are of two types. First, solutions of the form
$\{S^{II},S_n,S_n\}$, containing two narrow solitons $S_n$ within a wide soliton $S^{II}$. Second, we find solutions $\{S^{IV},S_n,S_m,S_m,S_n\}$, containing four solitons. Due to the symmetry of the system, we can have two pairs of different solitons embedded in the upper state of $S_{IV}$. As expected, cases with $n\neq m$ exist in areas of bistability of $P_n$ and $P_m$. In this figure, stable solutions are only partially shown to illustrate the complexity and richness of the hybrid system.

The search for real-time manipulation of frequency combs is a crucial property that is being explored in different situations \cite{he2019self,nazemosadat2021switching,talla2017existence,ivars2021reversible,ivars2023photonic}. In this work, we uncover a mechanism for a deterministic and reversible switch between states with different widths. Such control relies on the tilt of snaking structures of different solutions. This phenomenon is exemplified for solitons $S_n$ as provided in Fig.~\ref{fig:4}(a). The depicted numerical simulation is performed assuming the stable soliton $S_1$ as the initial condition. The detuning $\delta$ is adiabatically increased in time, up to the value of stability of $S_5$. Tracking the width of the LS, it is possible to observe the dynamic transition between them. Such transition is rapidly triggered when the different solitons reach their existence limit. Indeed, it is the shift of their existence intervals that allows for the reversible switching amongst different solitonic families. Panels $i)-v)$ in Fig.~\ref{fig:4} show the frequency combs associated to solitons $S_{1-5}$. 

Moreover, for normal GVD cavities, access to the homogeneous bistable area is essential to start the generation of frequency combs. Recently, excitation of dark solitons via self-injection-locking has been demonstrated \cite{wang2022self,jin2021hertz}. In the present work, we take advantage of the hybridisation of the system which displaces the existence region of some states outside the inaccessible uniform bistable area. {The recession of the first SN$_1$} introduces an access point to the homogeneous bistable area which can be excited by an adiabatic blue shift of the laser, as depicted in Fig.~\ref{fig:4}(b). The upper homogeneous solution is expected to evolve towards a $P_n$ rather than to the lower HS since the potential has an effect as the Turing instability. Notably, patterns exhibit a deterministic switching mechanism analogous to that observed in solitons. 

To conclude, we here present a novel scenario for the formation of localised structures. While it has been long-established that the building blocks for solitons in non-linear systems are states that are either patterns or flat states, we introduce a most unusual scenario in which the basal states (patterns / flat) acquire a dual behaviour so that the system may use one state as ’quasi-flat’ or as a pattern. Throughout the paper, such duality is discussed in the context of a driven dissipative cavity with normal GVD, described by the damped-driven NSE, under the introduction of a periodic non-Hermitian modulation. The reported results represent a general new paradigm to understand the formation of nonlinear localised structures, where physics from normal and anomalous GVD regimes blend. In particular, this unconventional hybridised scenario gives birth to an extraordinarily rich landscape where many different species of frequency combs coexist, being stable and accessible. Furthermore, the hybridisation of the LS formation scheme has important practical consequences for real-time reshaping of combs. Beyond the fundamental reported findings, these results pave the route for flexible frequency comb generation.

\begin{acknowledgments}
This work was supported by the Spanish government via PID2022-138321NB-C21 and Generalitat de Catalunya via 2021 SGR 00606; C.M. acknowledges PID2021-124618NB-C21, funded by MCIN/
AEI/10.13039/501100011033 and ‘ERDF: a way of mak-
ing Europe’ of the European Union, and Generalitat Valenciana via PROMETEO/2021/082.
\end{acknowledgments}

\bibliography{apssamp}

\end{document}